\documentclass[%
 reprint, superscriptaddress,
amsmath,amssymb,longbibliography,aps,prb]{revtex4-2}

\pdfoutput=1

\usepackage{graphicx}

\usepackage{float}
\usepackage{bm}
\usepackage{dcolumn}
\usepackage{hyperref}
\usepackage[english]{babel}
\usepackage{xcolor}

\usepackage{mathalpha}
\usepackage{amsmath,amsfonts}
\usepackage{booktabs}
\usepackage{comment}

\newcommand{\gap}{\Hat{\Delta}}

\newcommand{\bk}{\mathbf{k}}

\newcommand{\bv}{\mathbf{v}}
\newcommand{\br}{\mathbf{r}}
\newcommand{\bp}{\mathbf{p}}

\newcommand{\UTe}{UTe$_2$}

\begin{document}

\title{ Determination of gap structure of triplet superconductors from field-dependent Knight shift measurements}

\author{Ge Wang}
\affiliation{Department of Physics, University of Florida, Gainesville, FL 326011 USA}

\author{Andreas Kreisel}
\affiliation{Niels Bohr Institute, University of Copenhagen, Denmark }

\author{P.J. Hirschfeld}
\affiliation{Department of Physics, University of Florida, Gainesville, FL 326011 USA}

\date{\today}

\begin{abstract}
We analyze the spin susceptibility of spin-triplet superconductors from the zero-field to finite-field regimes, with emphasis on its implications for Knight-shift measurements. In the zero-field limit, we review the general expression for the static spin susceptibility and highlight the universal zero-temperature sum rule, $\sum_i \chi_{ii}(T\!=\!0)=2\chi^N$,
which constrains the residual susceptibility components for any triplet state. Using representative isotropic, helical, and chiral $\vec{d}$-vectors, we illustrate how the Knight shift encodes the spin configuration of the order parameter and show that the sum rule remains robust even for anisotropic Fermi surfaces. We then incorporate magnetic field effects through a semiclassical Doppler shift of quasiparticle energies in the vortex state. The resulting field dependence of
the susceptibility
—including both longitudinal (Knight-shift) and transverse magnetic susceptibility components—provides a sensitive probe of nodal directions and the momentum dependence of the $\vec{d}$-vector. Applying this framework to \UTe, we demonstrate how the distinct irreducible representations allowed by orthorhombic symmetry can be differentiated by their field-dependent susceptibility.
\end{abstract}

\maketitle

\section{Introduction}
\label{sec:intro}
The Knight shift experiment provides a powerful probe of the spin structure of the superconducting order parameter. In conventional spin-singlet superconductors, the spin susceptibility drops sharply below the critical temperature, leading to a large reduction in the Knight shift that reflects the formation of spin-singlet Cooper pairs \cite{Anderson1959}. By contrast, in spin-triplet superconductors, the spin susceptibility is expected to show less or no reduction, since some of  the paired electrons retain parallel spin alignment\cite{Brison2021,VollhardtWoelfle}. Moreover, in anisotropic triplet states, the Knight shift depends on the relative orientation between the applied magnetic field and the $\vec{d}$-vector: the drop in the Knight shift is minimal (or absent) when the field is perpendicular to $\vec{d}$, and becomes more pronounced when the field is parallel to $\vec{d}$. This directional sensitivity makes Knight shift measurements a crucial tool for distinguishing between singlet and triplet pairing states and for inferring the spin orientation of unconventional superconductors.   It was the lack of this expected directionality that famously provided an early clue to the problems with the identification of triplet superconductivity in Sr$_2$RuO$_4$ \cite{Annett2006,Pavarini2006,Pustogow2019}.

The uranium-based compound UTe$_2$ has rapidly emerged as one of the most intriguing candidates for spin-triplet superconductivity. Experimental evidence of triplet superconductivity includes an upper critical field that exceeds the Pauli limit in all directions, reentrant superconductivity, and power law temperature dependencies of thermal conductivity \cite{Ran2019,Metz2019}. Despite these signatures, the superconducting order parameter in UTe$_2$ remains unresolved, with different experiments supporting distinct odd-parity representations (Table~\ref{tab:irrep}): Josephson tunneling favors $B_{1u}$ \cite{b1u1}; Elastic-moduli and penetration-depth support $B_{2u}$ \cite{b2u1,Metz2019}; and STM studies on the (011) surface suggest compatibility with $B_{3u}$ \cite{b3u2}. Additional studies have proposed two-component states \cite{chiral1,chiral2}.

 In this context, Knight shift measurements are especially promising because of their sensitivity to the spin structure of Cooper pairs.
Recently, Knight shift measurements on UTe$_2$ have reported a decrease in all three crystallographic directions across the superconducting transition, though the magnitude of the drop is very small \cite{Aoki2023, Aoki2025}. At first glance, this behavior resembles the expectation for a spin-triplet superconductor: the reduction in spin susceptibility is far less pronounced than in a singlet state, and a finite residual susceptibility persists at low temperatures. However, the results are not fully consistent with simple theoretical expectations. In particular, theory imposes a zero-temperature sum rule requiring that the sum of the residual spin susceptibilities along the three axes equals two-thirds of the total normal-state value, $\sum_i\chi_{ii}(T\!=\!0)=\frac{2}{3}\sum_i\chi^N=2\chi^N$~\cite{Bernat2023}, a condition not clearly satisfied by the current experimental data if interpreted in this way.  Various  effects can modify or invalidate the sum rule for the spin susceptibility.  First, spin-orbit coupling (SOC) renders the susceptibility anisotropic, and  requires $\chi^N$ being replaced by $\chi^N_{ii}$.    Furthermore, the Knight shift  includes not only the shift proportional to the spin susceptibility, but also possible orbital and core polarization shifts, which can complicate interpretation. It is therefore unclear at present whether the Knight shift data on UTe$_2$ are consistent with any of the  proposed spin triplet states allowed in the orthorhombic crystal symmetry.

\begin{table}
    \centering
    \begin{tabular}{ccc}
    \toprule
    
         $\Gamma$ & Gap function $\Vec{d}(\bk)$     \\
         \midrule

         $B_{1u}$ &  ($p_1 k_y,~ p_2 k_x,~ p_3 k_x k_y k_z $)    \\
         $B_{2u}$ & ($p_1 k_z,~ p_2 k_x k_y k_z,~ p_3 k_x $)   \\
         $B_{3u}$ & ($p_1 k_x k_y k_z,~  p_2 k_z,~ p_3 k_y $)  \\
    
    \bottomrule
    \end{tabular}
    \caption{List of possible spin-triplet superconducting states for an orthorhombic crystal with strong spin orbit coupling. Here $p_{i=1,..,3}$ are constants and $\forall  p_i \in \mathbb{R}$.}
    \label{tab:irrep}
\end{table}

In this work, we examine how a finite magnetic field influences the Knight shift in a spin triplet state, potentially raising the sum rule value through the Doppler effect associated with supercurrents around vortices. While nonzero fields are often treated as a complication in interpreting Knight shift data, we find that they can in fact provide additional insight into the superconducting state. By analyzing how susceptibilities change with increasing magnetic field, we show that finite-field effects can be used to probe the pairing order parameter, offering information not only on  spin  but also orbital structure. We then apply this analysis to UTe$_2$ by considering candidate pairing states proposed for this material, demonstrating how finite-field spin susceptibilities can help discriminate between them.

The paper is structured as follows.  In Section I we first review the usual formalism for calculating the Knight shift in the superconducting state in the standard weak-field limit, and then include the effect of the orbital field by calculating the Doppler shift experienced by quasiparticles in the vortex lattice superflow field.  In Section II we present pedagogical results for various types of model triplet states over a spherical Fermi surface.  First, we derive and illustrate the sum rule on the susceptibility at $T=0$ and show how it is fulfilled for these model states.  Next, we present results for the field dependence of the anisotropic susceptibility for these states, and propose a protocol that can be used to determine the order parameter structure.  The full protocol assumes that magnetotropic susceptibility measurements are available\cite{Shekhter2023,Zambra2025}, but some information is obtainable even in their absence.  In Section III we present results for the simplest ``realistic" model for the Fermi surface of UTe$_2$, and compare with experimental data.

\section{Method}
\label{sec:method}
\subsection{Hamiltonian and spin susceptibility}
We start from a one-band mean-field Hamiltonian,
\begin{equation}
H = \sum_{\mathbf{k},\sigma} \xi_{\mathbf{k}} \,
c^{\dagger}_{\mathbf{k}\sigma} c_{\mathbf{k}\sigma}
+ \frac{1}{2} \sum_{\mathbf{k}}
\left[ c^{\dagger}_{\mathbf{k}\alpha} \,
\Delta_{\alpha\beta}(\mathbf{k}) \,
c^{\dagger}_{-\mathbf{k}\beta} + \text{H.c.} \right],
\end{equation}
where $\xi_{\mathbf{k}}$ is the band dispersion measured from the Fermi level. 
The triplet gap function is written in the standard form
\begin{equation}
\hat{\Delta}(\mathbf{k}) = 
i \big( \vec{d}(\mathbf{k}) \cdot \vec{\sigma} ) \sigma_y ,
\end{equation}
with Pauli matrices $\vec{\sigma}$ acting in spin space. We will consider only unitary states which have $\vec{d}\times \vec{d}^*=0$.
From this Hamiltonian, the Green’s function takes the form
\begin{equation}
    \check{G}(\bk,i\omega_\ell)=\frac{1}{(i\omega_\ell)^2-E_\bk^2}\begin{bmatrix}
    (i\omega_\ell+\xi)\sigma_0 & \gap_{\bk}\\
    \gap_{\bk}^\dagger & (i\omega_\ell-\xi)\sigma_0
\end{bmatrix} \label{eq:Green}
\end{equation}
in particle-hole and spin space and $\omega_\ell$ is a Fermionic Matsubara frequency.

The spin susceptibility is obtained from the Kubo formula,
\begin{equation}
\chi_{ij}(\mathbf{q}, i\Omega_n) 
= \int_0^{\beta} d\tau \, e^{i\Omega_n \tau} 
\langle T_\tau M_i(\mathbf{q},\tau) M_j(-\mathbf{q},0) \rangle ,
\end{equation}
where $M_i$ is the magnetization operator along $i$.
In this work we focus exclusively on the static, uniform spin susceptibility relevant for 
the Knight shift,
\begin{equation}
\chi_{ij} =
\lim_{\Omega \to 0} \chi_{ij}(\mathbf{q}=0, \Omega+i\delta).
\end{equation}

In the bare-bubble approximation (no vertex corrections), the spin susceptibility in the Matsubara representation is
\begin{align}
\chi_{ij}(i\Omega_n)
&= -\frac{1}{2\beta}\sum_{\mathbf{k},\,i\omega_\ell}
   \operatorname{Tr}\Bigg[
   \check{G}(\mathbf{k},i\omega_\ell)
   \begin{pmatrix}\sigma_i&0\\[2pt]0&-\sigma_i^{T}\end{pmatrix}
   \nonumber\\
&\qquad\qquad \times
   \check G(\mathbf{k},i\omega_\ell+i\Omega_n)
   \begin{pmatrix}\sigma_j&0\\[2pt]0&-\sigma_j^{T}\end{pmatrix}
   \Bigg] \label{eq:bubble},
\end{align}
and the physical susceptibility is $\chi(\Omega)=\chi(i\Omega_n)|
_{i\Omega_n\rightarrow \Omega+i\delta}$.
\begin{figure*}[t]
  \centering
      \includegraphics[width=\linewidth]{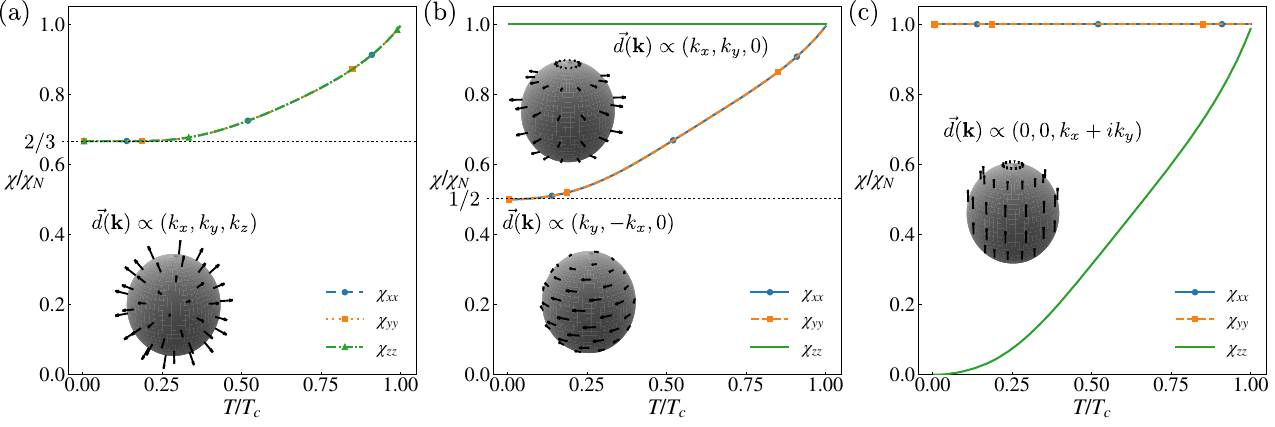}
  \caption{Spin susceptibilities for common triplet states: (a) Isotropic Balian-Werthamer state, (b)  two distinct helical states and (c) chiral state. The suppression of the spin susceptibility shows imprints of the direction of the $\vec d$-vector on the Fermi surface (insets), but in all cases the three components add up to $2\chi^N$ in the limit
  $T\rightarrow 0$.}
  \label{fig:chiT}
\end{figure*}

\subsection{Doppler shift and vortex averaging}

To incorporate the orbital effect of vortices in a magnetic field, we follow the semiclassical
approach introduced by Volovik for nodal superconductors \cite{Volovik1993,Kuebert1997,Vekhter1999}. 
In the presence of a supercurrent, the quasiparticle spectrum acquires a Doppler shift,
\begin{equation}
E_{\mathbf{k}} \;\to\; E_{\mathbf{k}} + E_D(\bk,\br,j),
\end{equation}
where \begin{equation}
    E_D(\bk,\br,j)\equiv \mathbf{p}_s(\br,j)\cdot\textbf{v}_F(\textbf{k})
\end{equation}
is the energy of Doppler shift, $\mathbf{p}_s(\br,j)$ is the superfluid momentum when magnetic field is parallel to $j$ direction, and $\textbf{v}_F(\bk)$ is the Fermi velocity.
Around a single vortex, the velocity decays as $|\mathbf{p}_s(\br,j)| \propto 1/r$
with distance $r$ from the vortex core.  
For a vortex lattice, we approximate the field-induced quasiparticle properties by
averaging physical quantities over a unit cell of area $\pi R^2$, where the intervortex
distance $R$ is set by the applied magnetic field through 
\begin{equation}
\pi R^2 = \frac{\Phi_0}{H},
\end{equation}
with $\Phi_0$ the flux quantum.  

Within this scheme, the Green's functions are modified by replacing $\omega$ 
with $\omega - E_D$, and the spin susceptibility is 
then evaluated by averaging the resulting expressions over the vortex unit cell (v.u.c.) by
\begin{equation}
    \Big\langle g(\br)\Big\rangle_\br\equiv \int_{\text{v.u.c.}}g(\br)d^2 r.
\end{equation}
This procedure captures the essential Doppler effect of circulating supercurrents 
and provides a tractable way to account for finite-field corrections to the Knight shift.

\section{Results}
\subsection{Spin susceptibility in zero field limit and sum rule}
We begin by reviewing the general form of the spin susceptibility  in the limit of an 
infinitesimal applied magnetic field, for readers not familiar with these basic properties of triplet superfluids. Starting from the Kubo formula introduced in Sec.~\ref{sec:method}, one finds that for a 
unitary triplet superconductor without SOC or with weak SOC the diagonal components of the static susceptibility tensor can be written as

\begin{equation}
\chi_{ii}(T) = \chi^N \Big\langle 
1 - \hat{d}_i^*(\mathbf{k}) \, \hat{d}_i(\mathbf{k}) \,
\big(1 - Y(\mathbf{k},T)\big)
\Big\rangle_{\text{FS}} ,
\label{eq:chi_general}
\end{equation}
where $\chi^N$ is the spin susceptibility of the normal state, 
\begin{equation}
    \Hat{d}_j(\bk)\equiv \frac{\Vec{d}(\bk)\cdot\Hat{e_j}}{|\Vec{d}(\bk)|}
\end{equation}
is the normalized 
$d$-vector, and $Y(\mathbf{k},T)$ is the Yosida function $Y({\bf k},T)=\int d\xi (-df/dE_{\bf k})$ describing the temperature 
dependence of quasiparticle excitations\cite{VollhardtWoelfle}. The angular brackets indicate a weighted average,
\begin{equation}
\langle f(\Omega_\bk) \rangle_{\text{FS}} \equiv 
\frac{\int d\Omega_\bk \, N_0(\Omega_\bk) \, f(\Omega_\bk)}
     {\int d\Omega_\bk \, N_0(\Omega_\bk)}.
\end{equation}
where $N_0(\Omega_\bk)=2|\mathbf \nabla \xi_{\mathbf k_F}|^{-1}$ is the local DOS on Fermi surface including both spins.

At zero temperature, $Y(\mathbf{k},0)=0$, so the residual spin susceptibility depends only 
on the spin orientation of the $d$-vector:
\begin{equation}
\chi_{ii}(0) = \chi^N \Big\langle 1 - \hat{d}_i^*(\mathbf{k}) \hat{d}_i(\mathbf{k}) \Big\rangle_{\text{FS}} .
\end{equation}
Summing over all directions gives
\begin{equation}
\chi_{xx}(0) + \chi_{yy}(0) + \chi_{zz}(0) = 2\,\chi^N,
\label{eq:sumrule}
\end{equation}
since $\sum_i \hat{d}_i^*(\mathbf{k}) \hat{d}_i(\mathbf{k})=| \hat{d}(\mathbf{k})|^2=1$.
This is the sum rule for triplet 
superconductors - the sum of residual spin susceptibilities equals $\frac{2}{3}$ the sum of normal state spin susceptibilities, $3\chi^N$, in the absence of spin-orbit coupling.

To illustrate the behavior of Knight shift in spin-triplet superconductors, we 
consider several representative $\vec{d}(\mathbf{k})$-vectors with spherically symmetric Fermi surface. These examples highlight 
the zero-temperature sum rule for the residual susceptibility and clarify that Knight shift 
is primarily sensitive to the spin configuration of the order parameter, rather than its orbital structure.  

\textbf{Isotropic state.}  
In the fully isotropic unitary triplet state\cite{Balian1963}, the $\vec{d}$-vector points equally along all spin 
directions, for example
\begin{equation}
\vec{d}(\mathbf{k}) \propto \hat{x} k_x + \hat{y} k_y + \hat{z} k_z .
\end{equation}

Here the spin degrees of freedom are equally shared among the three axes, leading to a uniform 
reduction of the spin susceptibility Fig.~\ref{fig:chiT} (a). At $T \to 0$, the residual susceptibility satisfies
\begin{equation}
\chi_{xx}(0) = \chi_{yy}(0) = \chi_{zz}(0) = \tfrac{2}{3} \chi^N ,
\end{equation}
which makes the sum rule most transparent.

\textbf{Helical state.}  
In a helical state  
\begin{equation}
\vec{d}(\mathbf{k}) \propto \hat{x} k_x \pm \hat{y} k_y {\rm ~or~} \hat{x} k_y \pm \hat{y} k_x,
\end{equation}
the spin susceptibility remains at its normal-state value for fields applied along $z$, while it is suppressed for 
fields applied along $x$ or $y$, see Fig.~\ref{fig:chiT}(b). Knight shift
experiments would therefore show a strong directional dependence, with one axis perpendicular to the $\vec d$-vector plane retaining
its normal-state value and the other two dropping when cooling below $T_c$.

\textbf{Chiral state.} Yet another common unitary triplet state is the
chiral $p$-wave, often written as
\begin{equation}
    \vec{d}(\mathbf{k}) \propto \hat{z}(k_x \pm i k_y).
\end{equation}
For ${\bf H} \parallel {\vec d}$, $\chi(0)\rightarrow 0$, while for  ${\bf H} \perp {\vec d}$, $\chi(0)=\chi
_N$, as expected but not observed for Sr$_2$RuO$_4$\cite{Annett2006}, see Fig.~\ref{fig:chiT}(c).

\begin{figure}[tb]
    \centering
    \includegraphics[width=\linewidth]{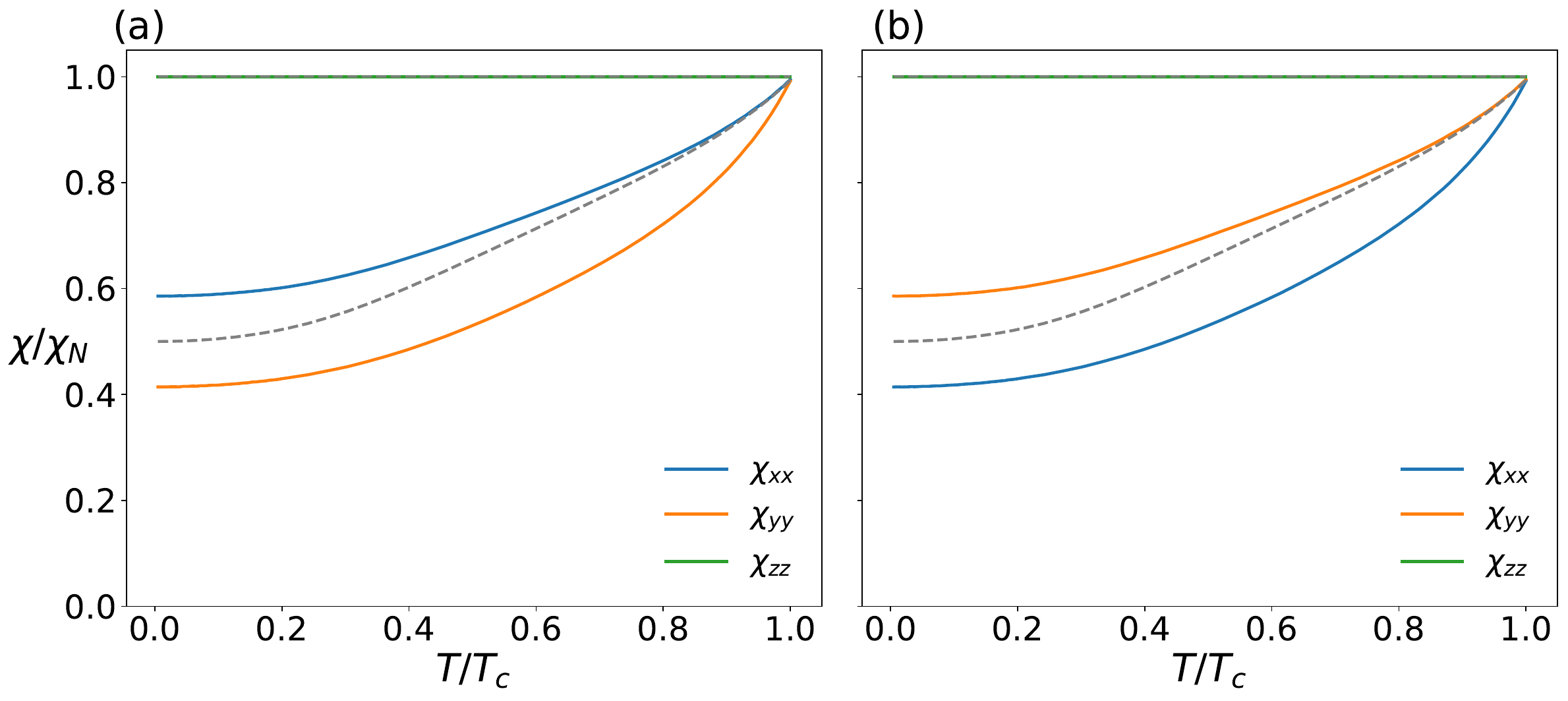}
    \caption{(a): $\vec{d}(\mathbf{k}) = \hat{x} k_y - \hat{y} k_x$ with anisotropic DOS
    $m_x=2m_y$.
  (b) $\vec{d}(\mathbf{k}) = \hat{x} k_x + \hat{y} k_y$ with anisotropic DOS
    $m_x=2m_y$.
  The dashed lines indicate isotropic case $m_x=m_y$.
  DOS anisotropy gives different susceptibility components for states with identical susceptibility components when isotropic, but the sum rule still holds.}
  \label{fig:aniso}
\end{figure}

\textbf{Effect of anisotropic density of states.}  
The spin susceptibility sum rule is a robust property of triplet superconductivity due to normalization of $\vec{d}(\bk)$ in the expression and remains valid even in the presence of 
anisotropy in the normal-state Fermi velocity. For example, an anisotropic density of states (DOS) arising 
from different effective masses $m_x, m_y, m_z$ can be treated 
by introducing dimensionless rescaled momenta
\begin{equation}
    \tilde{k}_i = \frac{k_i}{\sqrt{m_i}}, \qquad (i = x,y,z).
\end{equation}

Then the expression for spin susceptibility will be the same as isotropic case with modified angular weighting of the gap 
function $\hat{\Delta}(\bk')$. We consider two different helical states, $\vec{d}(\mathbf{k}) \propto \hat{x} k_y - \hat{y} k_x$  and $\vec{d}(\mathbf{k}) \propto \hat{x} k_x + \hat{y} k_y$ as example to show 2 key features of anisotropy:  first, it preserves the sum rule and second, it splits  susceptibility components degenerate over the isotropic Fermi surface. Over a spherical Fermi surface, these 2 states have identical $\chi(T)$ vs $T$. For an anisotropic Fermi DOS, one helical state is equivalent to an effective 
\begin{equation}
\vec{d}(\mathbf{k}) \propto \hat{x} \sqrt{m_y}k_y - \hat{y} \sqrt{m_x}k_x ,
\end{equation}
and the other state is equivalent to an effective
\begin{equation}
\vec{d}(\mathbf{k}) \propto \hat{x} \sqrt{m_x}k_x + \hat{y} \sqrt{m_y}k_y .
\end{equation}
The spin susceptibility components plotted in Fig.~\ref{fig:aniso} for the 2 states will no longer be the same, but the sum rule still holds,
\begin{equation}
    \sum_i \chi_{ii} = 2\chi^N
\end{equation} unless spin-orbit coupling is included, see below.  

\subsection{Spin susceptibility in low field}

The spin susceptibility in the presence of Doppler shift at $T=0$ is (see Appendix)
\begin{equation}
\chi_{ii}(T=0, H\!\parallel\!j)=\chi_{ii,j}^{\text{cond}}+\chi_{ii,j}^{\text{qp}} \label{eq:chiH}
\end{equation}
\begin{align}
\chi_{ii,j}^{\text{cond}}/\chi^N&=\Big\langle |\Hat{d}_{\perp}(\bk)|^2* \Theta(|\Vec{d}(\bk)|-|E_D(\bk,\br,j)|)\Big\rangle \label{eq:chiHcond}\\
    \chi_{ii,j}^{\text{qp}}/\chi^N
    &=\text{Re}{\Big \{ }\Big\langle (|\Hat{d}_i(\bk)|^2\frac{|E_D(\bk,\br,j)|}{\sqrt{|E_D(\bk,\br,j)|^2-|\Vec{d}(\bk)|^2}} \nonumber\\
    &+|\Hat{d}_\perp(\bk)|^2\frac{\sqrt{|E_D(\bk,\br,j)|^2-|\Vec{d}(\bk)|^2}}{|E_D(\bk,\br,j)|})\Big\rangle{\Big \}} \label{eq:chiHqp}
\end{align}
where 
\begin{equation}
    |\Hat{d}_\perp(\bk)|^2\equiv 1-|\Hat{d}_j(\bk)|^2,
\end{equation}
and $\Big\langle\dots\Big\rangle=\Big\langle\Big\langle\dots\Big\rangle_{\text{FS}}\Big\rangle_\br$ is average over the Fermi surface and over the vortex unit cell. In this formula, the field applied along $j$ direction dictates the superfluid momentum $\textbf{p}_s(\br,j)$, and susceptibility measured for $i$ direction picks out $\Hat{d}_i(\bk)$. The factor
$\displaystyle \frac{|E_D(\mathbf{k},r,j)|}{\sqrt{|E_D(\mathbf{k},r,j)|^2 - |\vec d(\mathbf{k})|^2}}$,
when averaged over the Fermi surface and the vortex unit cell, yields the
field-induced quasiparticle density of states associated with the Doppler
shift.

We distinguish between the longitudinal susceptibility, where $i\!=\!j$ and the transverse susceptibility, where $i\!\neq\! j$.  We emphasize that $\chi_{ii}(H \!\parallel\! j)=\partial M_i/\partial
H_i|_{H=H_j}$ with $i\!\neq\! j$, the susceptibility measured along $i$ under a
field applied along $j$,  is conceptually distinct from $\chi_{ij}$ with $i\!\neq\! j$, the off–diagonal
elements of the linear-response tensor $\chi_{ij}=\partial M_i/\partial
H_j|_{H\rightarrow 0}$.
We note that the transverse susceptibility
could, in principle, be accessed experimentally using magnetotropic 
susceptibility measurements \cite{Shekhter2023}. This technique measures the magnetic torque as 
the sample is rotated in a field and can directly probe the transverse 
magnetic response. In fact, our formulation of the transverse susceptibility closely 
parallels the concept introduced in Ref.~\cite{Zambra2025}.

\subsection{Using spin susceptibility to determine order parameter}
\begin{figure}[tbp]
     \includegraphics[width=\linewidth]{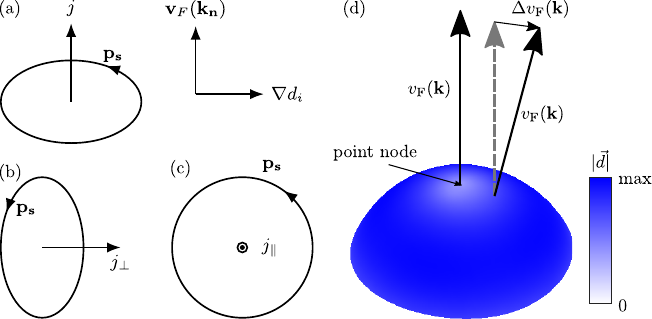}
     \caption{
Coordinates show direction of node and direction of increasing $d_i$ for (a)-(c). 
(a) field direction $j$ that is least affected by Doppler shift. 
(b) field direction $j_\perp$ where $|\nabla(\Delta\bv_F(\bk)\cdot \bp_s)\cdot \nabla d_i(\bk)|$ minimum. 
(c) field direction $j_\parallel$ where $|\nabla(\Delta\bv_F(\bk)\cdot \bp_s)\cdot \nabla d_i(\bk)|$ maximum. (d) Visualization of $\Delta v_F$ and gap magnitude around node.
  }
     \label{fig:config}
\end{figure}
In the presence of Doppler shift, the change of susceptibility with field can give more information on order parameter than magnitude of susceptibility at fixed field. The change in susceptibility along $i$ direction with a finite field in $j$ direction is given by 
\begin{widetext}
    \begin{align}
\Delta\chi_{ii}(H\!\parallel\! j)
&= \chi_{ii}(H\!\parallel\! j)-\chi_{ii}(H\rightarrow 0) \nonumber\\
&= \Big\langle 
\Theta(|E_D|-|\vec d|) \Big[
\underbrace{\frac{|E_D|}{\sqrt{|E_D|^2-|\vec d|^2}}}_{\textbf{positive}}
|\hat d_i|^2
+
\underbrace{\left(
\frac{\sqrt{|E_D|^2-|\vec d|^2}}{|E_D|}-1
\right)}_{\textbf{negative}}
|\hat d_\perp|^2
\Big] \Big\rangle \chi^N .
\label{eq:Dchi}
\end{align}

\end{widetext}

This expression shows first when $\Delta\chi_{ii}$ is positive, $d_i\neq 0$ because the prefactor of $|\Hat{d}_\perp(\bk)|^2$ is negative and the prefactor of $|\Hat{d}_i(\bk)|^2$ is positive.

For $\Delta\chi_{ii}\geq 0$, we can use $\Delta\chi_{ii}$ to determine the nodal direction (for point nodes) and $d_i$'s momentum dependence if transverse susceptibilities are also measured. The theta function in Eq.~\eqref{eq:Dchi} determines regions of the Fermi surface where the quasiparticle excitation energy vanishes due to the Doppler shift. Under the assumption that the Doppler energy scale
$E_D$ remains small compared to the superconducting gap magnitude, which is consistent with the 
low-field regime relevant for our analysis, these regions are in the vicinities of  nodes where $|\Vec{d}(\bk)|$ is small (and the standard nodal approximation is justified).

\begin{figure}[!htbp]
     \includegraphics[width=0.8\linewidth]{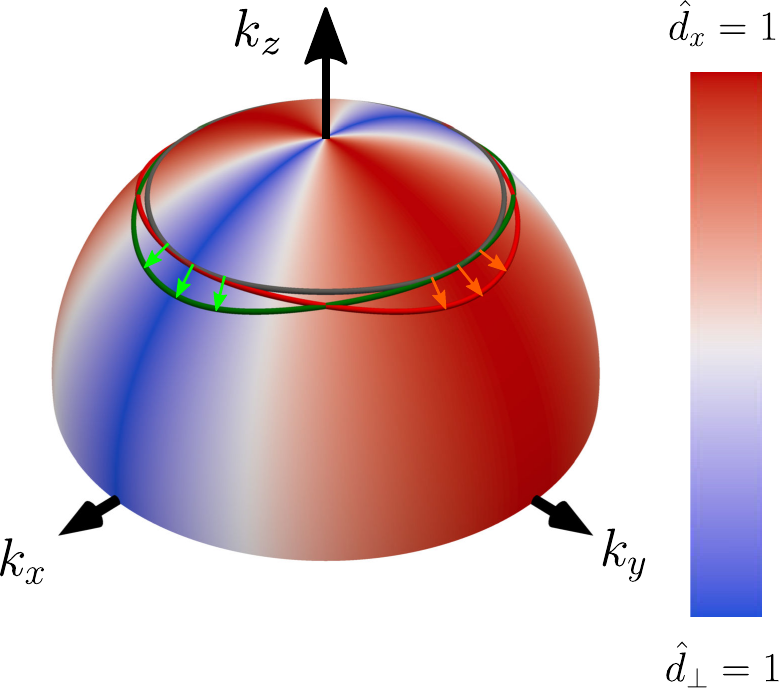}
     \caption{Quasiparticle excitation region (QP region) for example $B_{1u}$ state when field is perpendicular to nodal direction. Without $\Delta v_F$, the QP region is inside the gray boundary. With $\Delta v_F$, the QP region grows in perpendicular directions for $j_\perp$ (green) and $j_\parallel$ (red) field configurations. The heat map indicates the magnitude of $\Hat{d_i}$.}
     \label{fig:qp_region}
\end{figure}

If we fix $i$ and rotate the field ($j=x,y,z$), we will see the following pattern for $\Delta\chi_{ii}$. When $j\parallel\bk_n$ where $\bk_n$ is the nodal direction, the Doppler shift around nodes goes to 0 because $\textbf{p}_s$ ($\perp\bk_n$) and $\textbf{v}_F$ ($\parallel\bk_n$) are orthogonal (Fig.~\ref{fig:config}(a)). As a result, $\Delta\chi_{ii}(H\parallel\bk_n)\rightarrow 0$. On the other hand, when $j\perp \bk_n$, $\textbf{p}_s(\br,j)\cdot\textbf{v}_F(\textbf{k}_n)$ is maximized (with respect to field direction, see (b) and (c) of Fig.~\ref{fig:config}),  the theta function has support over a much larger region around the nodes. Thus we can expect a larger $\Delta\chi_{ii}$  (region inside the gray boundary in Fig.~\ref{fig:qp_region}) compared to $j\parallel\bk_n$ case.

\begin{figure*}[!htbp]
  \centering
\includegraphics[width=\linewidth]{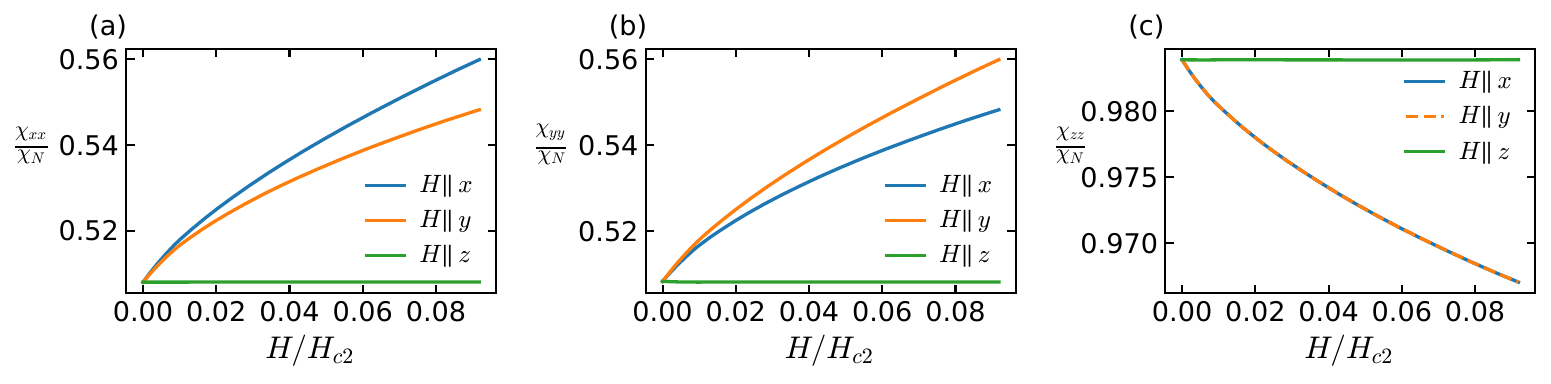}
    \caption{Field dependence of susceptibility for $B_{1u}$ state. When $H\rightarrow 0$, sum rule preserved but with finite field $\sum_i\chi_{ii}(H\!\parallel\!i)/\chi^N>2$. $\chi_{xx}$ and $\chi_{yy}$ both follow the field dependence $H\!\parallel\!\bk_n<H\!\parallel\! j_\perp\lesssim H\!\parallel\! j_\parallel$. $\Delta\chi_{zz}$ is negative because at first order $d_z=0$ for $B_{1u}$. Here the prefactors are set to $p_1=p_2=p_3=1$, but the qualitative trends of $\chi_{xx}$, $\chi_{yy}$, and $\chi_{zz}$ described above, which determine the gap structure, are insensitive to these prefactors. Although the 3 prefactors are equal, the $d_z=k_x k_y k_z$ has $|\hat{d_z}|^2$ integrates to orders of magnitude smaller over the Fermi surface than the other 2 directions and therefore $\chi_{zz}$ shows almost no reduction from normal state value at $T=0, H\rightarrow 0$.}
    \label{fig:chiH}
\end{figure*}

After the determination of nodal direction, a second order effect due to the variation of $\bv_F$ around a node allows one to determine the momentum dependence of $d_i$.
Let $\Delta\bv_F(\bk)$ be the correction to $\bv_F(\bk_n)$ near a node in the direction perpendicular to nodal direction, as in  Fig.~\ref{fig:config}(d). When $j$ rotates in the plane perpendicular to $\bk_n$, the susceptibility will grow more slowly with increasing field strength if the direction of the quasiparticle excitation region deformation is perpendicular to the direction of growing $d_i$ (call this direction $j_\perp$, see (b) of Fig.~\ref{fig:config}), i.e.
\begin{equation}
   \mathbf \nabla(\Delta\bv_F(\bk)\cdot \bp_s)\cdot \mathbf\nabla d_i(\bk)= 0, \label{perp direction}
\end{equation}
compared to when they are not perpendicular $\nabla(\Delta\bv_F(\bk)\cdot \bp_s)\cdot \nabla d_i(\bk)\neq 0$ because once the field direction meets the condition in Eq.~\eqref{perp direction}, the correction to $E_D$, $\Delta\bv_F(\bk)\cdot \bp_s$, is in the direction of $\Hat{d}_\perp(\bk)$, which gives an elongated quasiparticle excitation region along $\Hat{d}_\perp(\bk)$ (green arrows in Fig.~\ref{fig:qp_region}) and enhances the negative term of Eq.~\eqref{eq:Dchi}. When the field direction satisfies
\begin{equation}
    \nabla(\Delta\bv_F(\bk)\cdot \bp_s)\parallel \nabla d_i(\bk), \label{i direction}
\end{equation}
(call this direction $j_\parallel$, see Fig.~\ref{fig:config}(c)) the correction to $E_D$ is in the direction of growing $\Hat{d}_i$ , which gives an elongated quasiparticle excitation region along $\Hat{d}_i$ (red arrows in Fig.~\ref{fig:qp_region}) and enhances the positive term of Eq.~\eqref{eq:Dchi}. Thus $\Delta \chi_{ii}(H\parallel j_\parallel)$ will grow faster with increasing field than $\Delta \chi_{ii}(H\parallel j_\perp)$.

As an example, consider a $B_{1u}$ state of $D_{2h}$ as given in Table~\ref{tab:irrep} with spherical Fermi surface. The nodes are along the $z$ direction, so in Fig.~\ref{fig:chiH} when field is along $z$, susceptibility $\chi_{zz}(H\!\parallel\! z)$ is smallest. $
\Delta\bv_F(\bk) \propto k_x \hat{x} + k_y \hat{y}
$. $d_x(\bk)\propto k_y$ so $\mathbf\nabla d_x(\bk)\parallel y$. Now consider fields perpendicular to the nodal direction.  When $H\parallel x$, $\mathbf\nabla(\Delta\bv_F(\bk)\cdot \bp_s)\cdot \mathbf\nabla d_i(\bk)\neq 0$ and therefore the zero energy excitation has an extended support (red region in Fig.~\ref{fig:qp_region}) in $y$ direction compared to when field is along $y$ and $\nabla(\Delta\bv_F(\bk)\cdot \bp_s)\cdot\mathbf \nabla d_i(\bk)=0$. The latter has an extended zero-energy excitation support in the $x$ direction (green region in Fig.~\ref{fig:qp_region}). Larger zero-energy excitation support in the $\mathbf \nabla d_i(\bk)$ direction results in a faster growing susceptibility and thus we see in Fig.~\ref{fig:chiH} that $\chi_{zz}(H\!\parallel\! x)>\chi_{zz}(H\!\parallel\! y)$. Reversing this procedure allows us to find the direction of $\mathbf\nabla d_i(\bk)$ as it is parallel to the direction of field that has fastest susceptibility increase, and allows us to find the nodal direction as it is parallel to the direction of field that has slowest susceptibility increase.

In addition, we note the following features of the finite-field susceptibility shown in Fig.~\ref{fig:chiH}.
The zero-field sum rule is strictly preserved only in the limit $H \to 0$.
As the magnetic field increases, the total susceptibility sum
$\chi_{xx}(H\!\parallel \! x) + \chi_{yy}(H\!\parallel \! y) + \chi_{zz}(H\!\parallel \! z)$
rises monotonically toward its normal-state value, reflecting the field-induced quasiparticle population generated by Doppler shifts. The relative rates at which the individual components increase are independent of the prefactor coefficients of the $\vec{d}$-vector $(p_1, p_2$, and $p_3)$, and depend only on the irreducible representation of the order parameter.
\subsection{Competing field dependences}
For completeness, we note that the Zeeman effect has been neglected in the present analysis.
At low magnetic fields, the dominant contribution to the field dependence of the spin susceptibility
arises from the Doppler effect,
while the Zeeman splitting of quasiparticle bands provides only a subleading correction. If we compare the average Doppler shift energy $\langle E_D(\bf r)\rangle_r$ with the Zeeman energy $g\mu_B H$, it is easy to estimate that the Zeeman effect becomes dominant above a field $H^* \simeq (4\Delta^2/(\mu_B^2 g^2H_{c2}))$, of order 2T for UTe$_2$.  Thus the protocol discussed in this paper should be applied only below this crossover field.  

Another possible competing contribution to the field dependence of the spin susceptibility arises from vortex cores. Since the fraction of the sample occupied by vortex cores scales as $H/H_{c2,j}$ for a field applied along $j$ direction, the corresponding core contribution to $\Delta\chi_{ii}(H\parallel j)$ is expected to be approximately linear in field and can be estimated as
\begin{equation}
\Delta\chi^{\mathrm{core}}_{ii,j}(H)\approx \frac{H}{H_{c2,j}}\Big[\chi_{ii}(T=T_c)-\chi_{ii}(T=0)\Big],
\end{equation}
which reflects the approximate normal-state response of quasiparticles inside vortex cores.
This contribution may, in principle, be comparable in magnitude to the Doppler-induced quasiparticle contribution at low fields. However, it can be separated using the following calibration procedure: when the applied field is parallel to the nodal direction $j\!=\!\bk_n$, the Doppler contribution to $\Delta\chi_{ii}(H\!\parallel\! \bk_n)$ is suppressed (flat curves in Fig.~\ref{fig:chiH}), so the measured 
$\Delta\chi_{ii}(H\!\parallel\! \bk_n)$ is dominated by the core contribution $\Delta\chi^{\mathrm{core}}_{ii,\bk_n}(H)$, which can be identified experimentally by its approximately linear dependence on $H$, thereby providing an operational criterion for locating the nodal direction when it is not known a priori. Using the known anisotropy of $H_{c2,j}$, the core contribution for $j$ along other directions can then be obtained by
\begin{equation}
\Delta\chi^{\mathrm{core}}_{ii,j}(H)\approx \Delta\chi^{\mathrm{core}}_{ii,k_n}(H)\,\frac{H_{c2,k_n}}{H_{c2,j}},
\end{equation}
and subtracted from the corresponding data to isolate the Doppler contribution.

\section{Application to UT\protect\lowercase{e}$_2$}
\label{sec:realistic}

\begin{figure}[tb]
     \includegraphics[width=0.5\linewidth]{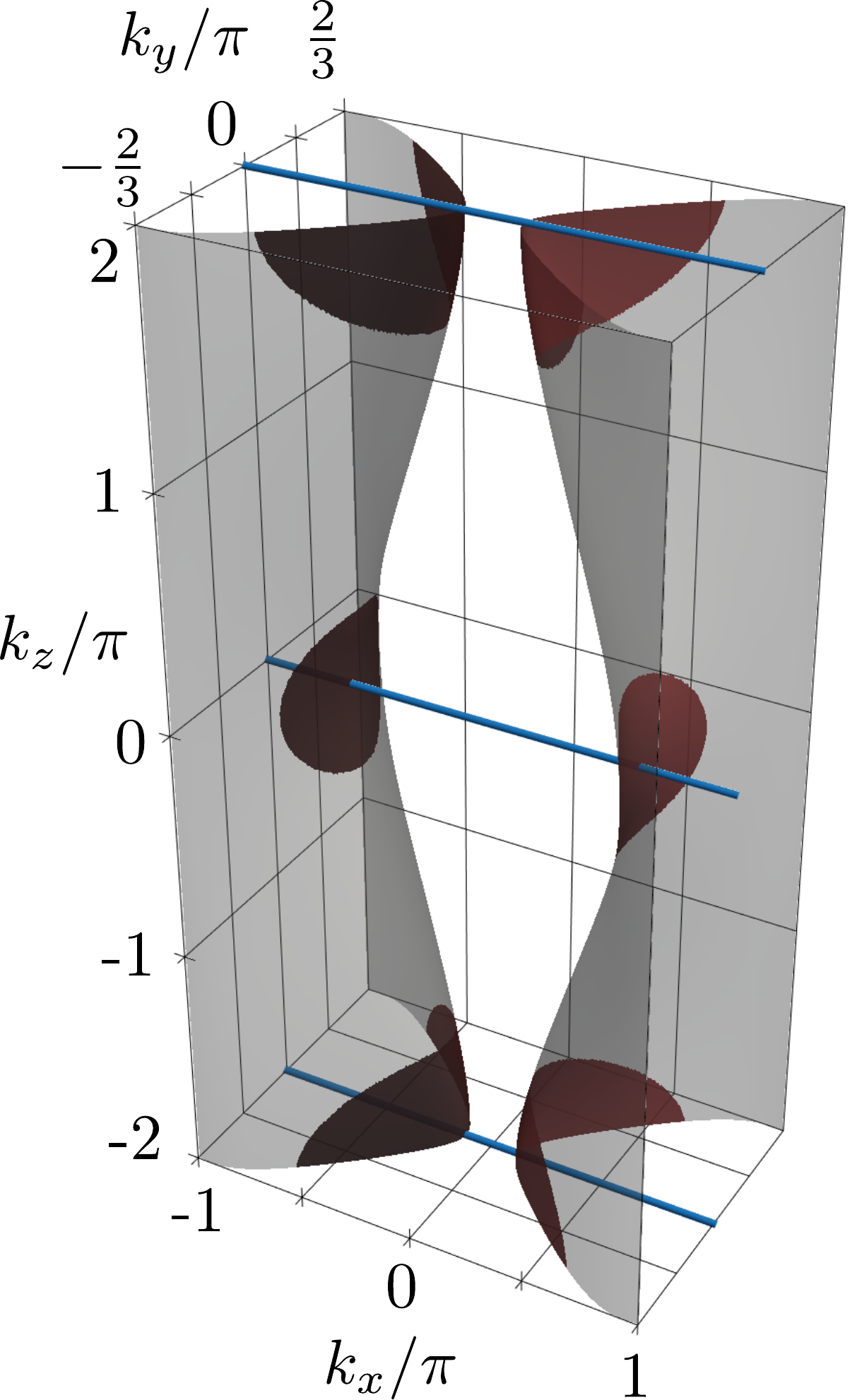}
     \caption{The Fermi surface of UTe$_2$ model. Plotted on the Fermi surface are nodal lines $k_y=0$, $k_z=\pm 2\pi$ and $k_y=0$, $k_z=\pm 2\pi$ for $B_{2u}$ state, and quasiparticle excitation region near the nodes due to the magnetic field in $y$ direction.}
     \label{fig:qpb3u}
\end{figure}

Having established the field dependent susceptibility as a probe for triplet superconductor order parameters, we now apply this method to the case of UTe$_2$. To capture the key symmetry and Fermi-surface features of UTe$_2$, we adopt the tight-binding Hamiltonian proposed by Shishidou \textit{et al.} \cite{Shishidou2021}. The model originates from DFT+U calculations that revealed a topological electron band near the chemical potential, driven by band inversion between even- and odd-parity combinations of the two U atoms forming a ladder-rung sublattice. In the normal state, the Hamiltonian takes the form
\begin{align}
H_N &= \epsilon_0(\mathbf{k}) - \mu 
   + f_{A_g}(\mathbf{k})\,\tau_x 
   + f_z(\mathbf{k})\,\tau_y
   + f_y(\mathbf{k})\,\sigma_x\tau_z \notag\\
   &\quad 
   + f_x(\mathbf{k})\,\sigma_y\tau_z
   + f_{A_u}(\mathbf{k})\,\sigma_z\tau_z ,
\end{align}
with 
\begin{align}
\epsilon_0(\mathbf k) &= t_1 \cos(k_x) + t_2 \cos(k_y) \nonumber\\
f_{Ag}(\mathbf k) &= m_0 + t_3 \cos(k_x/2) \cos(k_y/2) \cos(k_z/2) \nonumber\\
f_z((\mathbf k) &= t_z \sin(k_z/2) \cos(k_x/2) \cos(k_y/2) \nonumber\\
f_y(\mathbf k) &= t_y \sin(k_y) \nonumber\\
f_x(\mathbf k) &= t_x \sin(k_x) \nonumber\\
f_{Au}(\mathbf k) &= t_u \sin(k_x/2) \sin(k_y/2) \sin(k_z/2)
\end{align}
 where $\tau_i$ are the Pauli matrices acting in orbital space spanned by the two U orbitals and $\sigma_i$ are the Pauli matrices acting in spin space. The functions ($f_{Ag}$, $f_z$) give the dominant hybridization between two U atoms and the functions ($f_x$, $f_y$, $f_{Au}$) give the spin orbit coupling. We use the parameters $(\mu, t_1, t_2, m_0, t_3, t_z, t_x, t_y, t_u) = $ (0.129, -0.0892,\,0.0678, -0.062, 0.0742, -0.0742, 0.006, 0.008, 0.01) given in Ref.~\cite{Shishidou2021}.   The model yields a pair of quasi-cylindrical Fermi surface sheets open along $k_z$ direction Fig.~\ref{fig:qpb3u}. It reproduces the electron-like Fermi pocket observed in quantum oscillation experiment \cite{Ran2023, Eaton2024} and further supported by ARPES measurements\cite{Wray2020}, but not the orthogonal hole-like pocket or the controversial pocket on the $c$-axis face of the Brillouin zone\cite{Ran2023,Wray2020}. 

 This Fermi surface serves as the primary input for the present susceptibility analysis, which relies on Fermi surface averages. The Fermi surface
supports a symmetry-enforced nodal structure only for the 
$B_{3u}$ order parameter; other representations $B_{1u}$ and $B_{2u}$ do not possess symmetry-protected nodes on this particular 
Fermi surface. 
In contrast, a Fermi surface composed of two pairs of quasi-cylindrical sheets 
oriented along the orthogonal $k_x$ and $k_y$ directions, as suggested 
by quantum-oscillation studies, could in principle support both 
$B_{3u}$ and $B_{2u}$ nodal structures. 
Accordingly, in what follows we present the field-dependent 
susceptibility obtained from our method using the Hamiltonian for the $B_{3u}$ state; the behavior for $B_{2u}$, if 
such a configuration were realized, follows analogously and will be 
discussed briefly later.

The $B_{3u}$ basis functions are odd under
$k_y\!\to\!-k_y$ and $k_z\!\to\!-k_z$, giving rise to symmetry-enforced
nodes wherever the Fermi surface intersects the planes $k_y=0$ and
$k_z=0$ or $k_z=2\pi$ (due to the doubled
periodicity associated with the body-centered orthorhombic structure of the crystal). 
The resulting nodal points occur near
\[
\pm\bk_{n,1} = (\pm k_1,0,0) 
\quad\text{and}\quad
\pm\bk_{n,2} = (\pm k_2,0,2\pi),
\]
where $k_{1}$ and $k_{2}$ are constants set by the Fermi surface condition.
The gap near each pair of nodes can thus be expanded independently as
\begin{align}
\vec{d}(\bk=\pm\mathbf{k}_{n,1}+\mathbf{q})
&\approx a_{1}q_{z}\Hat{y} + b_{1}q_{y}\Hat{z},\\[2pt]
\vec{d}(\bk=\pm\mathbf{k}_{n,2}+\mathbf{q})
&\approx a_{2}q_{z}\Hat{y} + b_{2}q_{y}\Hat{z}
\end{align}
The constant coefficients $a_i,b_i$ are obtained from the first-order expansion of the gap function near each node; while their values depend on the exact order parameter, they do not affect the qualitative behavior of $\Delta\chi(H)$, which is enough for order parameter determination in our method.
The field-induced change in susceptibility $\Delta\chi_{ii}(H\parallel j)$ then depends only on regions around $\pm\bk_{n,1}, \pm\bk_{n,1}$  because of $\Theta(|E_D(\bk,\br,j)|-|\Vec{d}(\bk)|)$ in Eq.~\eqref{eq:Dchi}. The nonvanishing regions of the theta function (i.e. the quasiparticle excitation regions) for this model are shown in Fig.~\ref{fig:qpb3u}.

For UTe$_2$ the coherence lengths, $H_{c2}$, and vortex lattice relevant to vortex averaging are 
anisotropic. In the present work we use an angle-averaged (effectively 
isotropic) treatment of the Doppler energy. Importantly, the anisotropy does 
not alter the ordering of the field-induced changes 
$\Delta\chi_{ii}(H\!\parallel j)$ that distinguishes the odd-parity 
representations. Our conclusions therefore do not depend on the detailed 
vortex structure or critical field anisotropy.

\begin{figure*}[tb]
  \centering
\includegraphics[width=\linewidth]{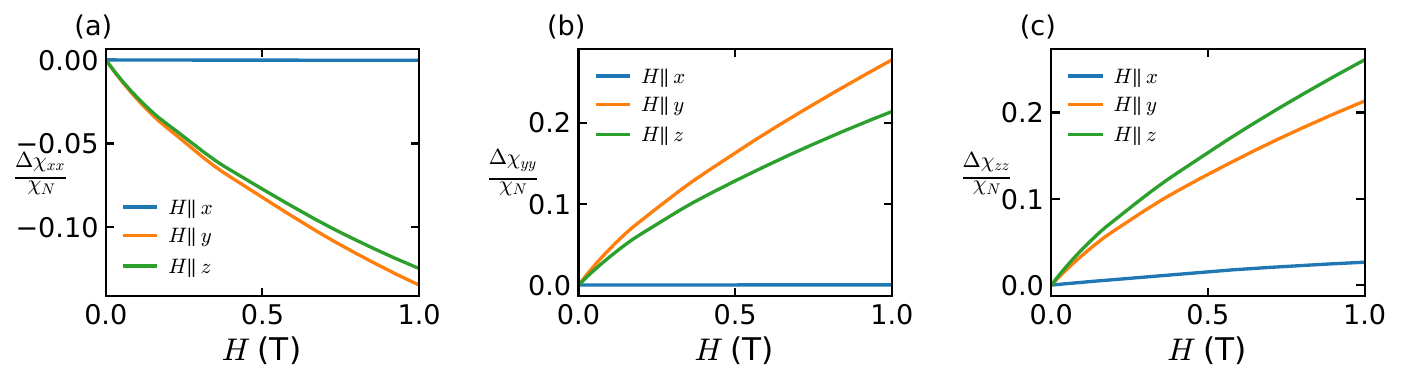}
    \caption{$\Delta\chi(H)$ for $B_{3u}$ for the three different directions (a-c) with Fermi surface model relevant for UTe$_2$. Quasiparticle excitation has a larger percentage effect on cylindrically shaped Fermi surface than on spherical Fermi surface (results shown in Fig.~\ref{fig:chiH}).}
    \label{chiHB3u}
\end{figure*}

The numerical calculation of $\Delta\chi(H)$ for $B_{3u}$ gives
\begin{align}
    &\Delta\chi_{xx}(H\!\parallel\! y) \approx \Delta\chi_{xx}(H\!\parallel\! z) \approx \Delta\chi_{xx}(H\!\parallel\! x)\lesssim 0\nonumber\\
    &\Delta\chi_{yy}(H\!\parallel\! y) > \Delta\chi_{yy}(H\!\parallel\! z) > \Delta\chi_{yy}(H\!\parallel\! x)\geq 0\nonumber\\    &\Delta\chi_{zz}(H\!\parallel\! z) > \Delta\chi_{zz}(H\!\parallel\! y) > \Delta\chi_{zz}(H\!\parallel\! x)\geq 0,
\end{align}
as shown in Fig. \ref{chiHB3u}. This result is plotted with $a_n=1,b_n=1$, $n=1,2$, but as mentioned in the previous section, the ordering does not depend on the values $a_n,b_n$, which ultimately are derived from the coefficients $p_i$ in the $\Vec{d}$-vector, cf. Table \ref{tab:irrep}.

This result can be interpreted
by noting that the Fermi surface integral of Eq.~\eqref{eq:Dchi} can be evaluated by summing the contributions from each nodal point,
\begin{equation}
    \Delta\chi_{ii}(H\!\parallel\! j)=\sum_n \Delta\chi_{ii,n}(H\!\parallel\! j),
\end{equation}
where $n$ labels nodes, and each node yields a consistent
anisotropy pattern in $\Delta\chi_{ii,n}$ according to the discussion above, namely, $\Delta\chi_{ii,n}(H\!\parallel\!\bk_n)<\Delta\chi_{ii,n}(H\!\parallel\! j_\perp)<\Delta\chi_{ii,n}(H\!\parallel\! j_\parallel)$.
This anisotropy provides a clear 
signature for the $B_{3u}$ state. In contrast, a $B_{2u}$ order parameter 
would exhibit a different pattern given in Table~\ref{tab:dchi_ordering}.
Such distinct field-orientation trends thus allow a direct experimental 
identification of the pairing symmetry.

\begin{table*}[!htbp]
\centering

\begin{tabular}{l|c|c|c}

\toprule
 & $B_{1u}$ & $B_{2u}$ & $B_{3u}$ \\
\midrule
$\Delta\chi_{xx}$ & $H\!\parallel\!x>H\!\parallel\!y>\,H\!\parallel\!z\geq 0$ 
               & $H\!\parallel\!x>H\!\parallel\!z>\,H\!\parallel\!y\geq 0$
               & $H\!\parallel\!x\approx H\!\parallel\!y\approx \,H\!\parallel\!z\lesssim 0$ \\[4pt]
$\Delta\chi_{yy}$ & $H\!\parallel\!y>H\!\parallel\!x>\,H\!\parallel\!z\geq 0$ 
               & $H\!\parallel\!x\approx H\!\parallel\!y\approx \,H\!\parallel\!z\lesssim 0$ 
               & $H\!\parallel\!y>H\!\parallel\!z>\,H\!\parallel\!x\geq 0$ \\[4pt]
$\Delta\chi_{zz}$ & $H\!\parallel\!x\approx H\!\parallel\!y\approx \,H\!\parallel\!z\lesssim 0$ 
               & $H\!\parallel\!z>H\!\parallel\!x>\,H\!\parallel\!y\geq 0$ 
               & $H\!\parallel\!z>H\!\parallel\!y>\,H\!\parallel\!x\geq 0$ \\

\bottomrule

\end{tabular}
\caption{Field-direction ordering of $\Delta\chi_{ii}(H)$ components for the three odd-parity representations. 
Each entry lists the relative magnitude of $\Delta\chi_{ii}$ under different field orientations (see text). The $B_{1u}$ case is only relevant if a Fermi surface pocket along c-axis is present, as suggested in some experiments\cite{Ran2023,Wray2020}. The $B_{2u}$ state would require electronic structure that consists of four tubes of Fermi surfaces\cite{Ran2023, Eaton2024,Wray2020}.}
\label{tab:dchi_ordering}
\end{table*}

\section{Discussion}
\label{sec:discussion}

In this work we have shown that the field dependence of the Knight shift provides
information beyond the zero-field spin susceptibility and offers a powerful
means of distinguishing the odd-parity pairing states allowed by orthorhombic
symmetry. The central result is the characteristic ordering of the finite-field
susceptibility changes $\Delta\chi_{ii}(H\!\parallel j)$, which depends only on the
symmetry of the superconducting order parameter. The Doppler shift generated by
vortex-induced superflow produces a strongly anisotropic excitation pattern near
the gap nodes, allowing one to infer both the nodal direction and the momentum
dependence of the corresponding $d$-vector component $d_i(\bk)$. Applying this
framework to the minimal tight-binding description of UTe$_2$, we illustrate how
the $B_{3u}$ state yields a distinctive hierarchy among the field-induced
susceptibility changes.

The tight-binding model from Ref.~\cite{Shishidou2021} employed here captures key symmetry
features, and the quasi-cylindrical Fermi-surface topology
associated with the U  states. However, this minimal two-orbital model, even with SOC, does
not reproduce the large anisotropy of the normal-state magnetic susceptibility
observed experimentally for UTe$_2$. This limitation reflects its purpose as a low-energy
model rather than a quantitative description of all contributions to
the magnetic response. In particular, a realistic account of the susceptibility anisotropy would require anisotropic $g$-factors. More generally, extending the model to include Te-orbital character and correlation effects may improve quantitative details, but is not expected to alter the symmetry-controlled ordering of the field-induced susceptibility between different field directions.

If UTe$_2$ possesses additional Fermi-surface cylinders, such as the orthogonal
quasi-cylindrical sheets suggested in several quantum oscillation
experiments, then a $B_{2u}$ order parameter could develop nodal intersections on
those surfaces, since its symmetry-enforced nodal lines $k_x = 0$ and
$k_z = 0$ would naturally intersect such sheets.  
In that case, if the superconducting state were $B_{2u}$, the resulting
field-induced susceptibility anisotropy $\Delta\chi_i(H\!\parallel j)$ would be
expected to follow the qualitative behavior listed in the $B_{2u}$ column of
Table~\ref{tab:dchi_ordering}.
Similarly, the $B_{1u}$ case is only relevant if a Fermi surface pocket along c-axis is present, as suggested in some experiments\cite{Ran2023,Wray2020}.

An additional advantage of the present approach is that it
continues to provide directional information even when the
nodal structure deviates from the high–symmetry axes
assumed for the minimal odd-parity representations. Several proposed nonunitary states of UTe$_2$ possess point nodes that are displaced from a crystallographic axis~\cite{chiral2}.
In addition, recent work has shown that even within a single component $B_{2u}$ or $B_{3u}$ state, additional nodes away from the high symmetry directions emerge within a model including U- and Te-derived bands~\cite{Christiansen2025}. 
In such cases, if the
point nodes reside in (or close to) a plane, the field-induced
susceptibility change $\Delta\chi_{ii}(H\!\parallel\! j)$ is expected to
grow most rapidly when the applied field is oriented
perpendicular to that plane, since the Doppler shift most
efficiently excites quasiparticles in the nodal region. When
the field lies within the plane, and the nodes are only
slightly offset from a high-symmetry axis, the increase of
$\Delta\chi_{ii}$ with field strength will be slowest when the
field is aligned with that axis—thus preserving a robust
signature of the underlying nodal orientation. By contrast,
if the nodal structure is genuinely three-dimensional and not
confined to an approximate axis or plane, the anisotropy of
the finite-field response is expected to be weak, and the
method may no longer reliably determine either the nodal
direction or the momentum dependence of $d_i(\mathbf{k})$.

For Knight-shift experiments, although the finite-field effect discussed here
provides a possible explanation for the small reductions observed in all field
directions, there are other effects that can cause the measured Knight shift to
deviate from the ideal spin-susceptibility sum rule. Orbital effects,
diamagnetic shielding currents, and core polarization all modify the local magnetic field at
the nuclear site. These contributions are not expected to exhibit strong field
dependence, so the field-induced change of the susceptibility may provide
a more reliable probe of the superconducting spin response than the absolute
magnitude of the Knight shift.

\section{Conclusion}
The identification of the order parameter symmetry in the low-field phase of UTe$_2$ has proven to be elusive, but the task is important to establish an anchor to eventually understand the rich and fascinating phase diagram in temperature, magnetic field, and pressure.
In this work, we have proposed that the field-induced susceptibility anisotropy provides a robust and
symmetry-sensitive probe of odd-parity superconductivity. The
ordering of the transverse susceptibilities $\Delta\chi_{ii}(H\!\parallel j)$ predicted here is controlled solely
by the structure of the $\vec d$-vector and remains stable against reasonable
variations in band structure, vortex anisotropy, and normal-state factors. Combined Knight shift and magnetotropic susceptibility
measurements at low fields should offer a route to identifying the
superconducting order parameter in UTe$_2$.


\begin{acknowledgments}
We acknowledge valuable conversations with J.-P. Brison, S.E.  Brown, K.A. Modic,  V. Mishra, and B.J. Ramshaw.   P.J.H. and G.W. were partially supported by
NSF-DMR-2231821. A.K. acknowledges support by the
Danish National Committee for Research Infrastructure
(NUFI) through the ESS-Lighthouse Q-MAT.
\end{acknowledgments}
\appendix
\section{Derivation of Eq.~(\ref{eq:chi_general})}

Starting with Eq.~(\ref{eq:bubble}) and plugging in the Green's function Eq.~(\ref{eq:Green}), we get
\begin{equation}
    \chi_{ii}=-\frac{1}{2\beta}\sum_{\bk,i\omega_l}
\frac{(i\omega_l)^2+\xi^2-|\vec{d}(\bk)|^2+2|d_i(\bk)|^2}{((i\omega_l)^2-E_\bk^2)^2}.
\end{equation}

Here we assume vanishing or weak SOC, such that spin remains approximately a
good quantum number in the band basis and SOC-induced corrections to the spin
operator are negligible.

Performing the Matsubara sum, we obtain
\begin{align}
    &\chi_{ii}=-\frac{1}{2}\sum_{\bk}\nonumber \\
    &\Bigl[\bigl(-2\frac{\beta/4}{\cosh^2(\beta E_\bk/2)}+\frac{\tanh(\beta E_\bk/2)}{E_\bk}\bigr)\bigl(1-\frac{\xi^2}{E_\bk^2}\bigr)(2|\Hat{d}_j|^2-1) \nonumber \\
    &-2\frac{\beta/4}{\cosh^2(\beta E_\bk/2)}-\frac{\tanh(\beta E_\bk/2)}{E_\bk}\nonumber \\
    &+\frac{\xi^2}{E_\bk^2}\bigl(-2\frac{\beta/4}{\cosh^2(\beta E_\bk/2)}+\frac{\tanh(\beta E_\bk/2)}{E_\bk}\bigr)\Bigr].
\end{align}
The terms in summand have either a $\displaystyle \frac{\partial f(E_\bk)}{\partial E_\bk}$ or $\displaystyle\frac{d}{d\xi}(\frac{\xi}{E_\bk})$ factor, which sharply peaks around the Fermi surface. Therefore, we can approximate
\begin{equation}
    \sum_\bk \approx \int d\Omega_\bk N_0(\Omega_\bk)\int d\xi.\label{xi}
\end{equation}
Using 
\begin{equation}
    \int d\xi \frac{d}{d\xi}(\frac{\xi}{E_\bk})\approx 2
\end{equation}
and the definition
\begin{equation}
    Y(\bk,T)=-\int d\xi \frac{\partial f(E_\bk)}{\partial E_\bk},
\end{equation}
we get the result Eq.~(\ref{eq:chi_general}).

\begin{widetext}
\section{Derivation of Eqs.~(\ref{eq:chiH}) to~(\ref{eq:chiHqp})}

When there is Doppler shift, we replace the $\omega$ in Green's function Eq.~(\ref{eq:Green}) with $\omega-E_D(\bk,\br,j)$ and plug into susceptibility formula Eq.~(\ref{eq:bubble}). We get temperature dependent susceptibility in superconducting state
\begin{align}
\chi_{ii}(T, H\!\parallel\!j)
&= \Bigg\langle-\frac{1}{2}\sum_{\bk}\nonumber \\
&\Bigg[
\Big(
   -\frac{\beta/4}{\cosh^2\!\big[\beta (E_\bk+E_D)/2\big]}
   -\frac{\beta/4}{\cosh^2\!\big[\beta (E_\bk-E_D)/2\big]}
   -\frac{f(E_\bk+E_D)}{E_\bk}
   +\frac{f(-E_\bk+E_D)}{E_\bk}
 \Big)
 \Big(1-\frac{\xi^2}{E_\bk^2}\Big)(2|\hat d_i|^2-1)
\nonumber\\[4pt]
&\qquad
-\frac{\beta/4}{\cosh^2\!\big[\beta (E_\bk+E_D)/2\big]}
-\frac{\beta/4}{\cosh^2\!\big[\beta (E_\bk-E_D)/2\big]}
+\frac{f(E_\bk+E_D)}{E_\bk}
-\frac{f(-E_\bk+E_D)}{E_\bk}
\nonumber\\[4pt]
&\qquad
+\Big(
   -\frac{\beta/4}{\cosh^2\!\big[\beta (E_\bk+E_D)/2\big]}
   -\frac{\beta/4}{\cosh^2\!\big[\beta (E_\bk-E_D)/2\big]}
   -\frac{f(E_\bk+E_D)}{E_\bk}
   +\frac{f(-E_\bk+E_D)}{E_\bk}
 \Big)\frac{\xi^2}{E_\bk^2}
\Bigg]\Bigg\rangle_\br .
\label{eq:B1}
\end{align}
At zero temperature, the expression simplifies to the sum of 
\begin{align}
    \chi_{ii,j}^\text{cond}=\Big\langle\sum_{\bk}&
\Big(\frac{\Theta (E_\bk-E_D)}{E_\bk}-\frac{\Theta (-E_\bk-E_D)}{E_\bk}\Big)
\bigl(1-\frac{\xi^2}{E_\bk^2}\bigr )\bigl(1-|\Hat{d}_i|^2\bigr)
\Big\rangle_\br
\end{align}
and 
\begin{equation}
    \chi_{ii,j}^\text{qp}=\Big\langle\sum_{\bk}
\Big[\delta(E_\bk+E_D)+\delta(-E_\bk+E_D)\Big]\Big(|\Hat{d}_i|^2-|\Hat{d}_i|^2\frac{\xi^2}{E_\bk^2}+\frac{\xi^2}{E_\bk^2}\Big)\Big\rangle_\br.
\end{equation}
Integrating over $\xi$ as in Eq.~(\ref{xi}) we get Eq.~(\ref{eq:chiHcond}) and Eq.~(\ref{eq:chiHqp}).

\end{widetext}

\bibliography{bibliography}

\end{document}